# Magnetocaloric effect and nature of magnetic transition in low dimensional DyCu$_2$

M. Venkat Narayana[1] Ganesh Kotnana and S. Narayana Jammalamadaka[1*]

[1]*Magnetic Materials and Device Physics Laboratory, Department of Physics, Indian Institute of Technology Hyderabad, Hyderabad, India – 502 285.*

*Corresponding author: surya@iith.ac.in*

**Abstract:** In this manuscript, we propose a method to prepare small flakes of DyCu$_2$. On top of that we also report on the magnetocaloric effect and nature of magnetic transition of a strongly anisotropic DyCu$_2$ in its low dimension. Magnetization measurements were carried out in the temperature range of 5 – 100 K and up to the maximum magnetic field strength of 50 kOe. Magnetic entropy change ($\Delta S_M$) is estimated using the well-known Maxwell's equations and is found to be - 4.31 J/kg – K. Indeed, the $\Delta S_M$ peak broadened marginally compared with its bulk DyCu$_2$ and such a broadening can be attributed to significant increase in the total grain boundary volume. As these small flakes consists larger $\Delta S_M$ values at temperatures higher than the Nèel temperature ($T_N$), one can use them as a magnetic refrigerant material in a broad temperature range. We also plotted the M$^2$ vs. H/M (which are called as the Arrott plots) in order to find the nature of magnetic transition. Arrott plots infer that indeed there exists nonlinearity in M$^2$ vs. H/M behavior and such nonlinear behavior is ascribed to the random anisotropy or a random field that is present in the system.

Keywords: Magnetocaloric effect, Arrott plots, magnetic refrigeration and entropy change

PACS: 75.30.Sg, 75.75.-c, 75.20.En, 75.30.Cr, 75.30.Kz and 61.05.C-

**Introduction**

Advancement in energy saving technology would indeed help in developing the next generation energy storage devices. Among such technologies, technology based on the magnetic refrigeration has been a promising one and works based on the magnetocaloric effect (MCE) which essentially furnishes an alternative environmental friendly method if we compare with other cooling techniques. Basically MCE is a phenomenon of change in the temperature (heating or cooling) of a magnetic material upon the application of the magnetic field. Such phenomenon can be realized indirectly as a change in magnetic entropy ($\Delta S_M$)[1-4]. Materials which exhibits the MCE has gained much interest due to their high efficiency and non pollution which are the added advantages of the magnetic refrigeration technology compared with the gas refrigeration. In order to realize advancement in the technology, there is a quest for the development of the new materials. Many researchers have been developing new materials and during the course of search for the new materials, various authors have reported on the magnetic refrigeration materials such as $Gd_{0.7}Dy_{0.27}$[5], La-Fe-Si[6], $Tb_5Si_3$[7], and $Nd_6Co_{1.67}Si_3$[8] and $DyCu_2$[9].

Interestingly, magnetic properties of an orthorhombic $RCu_2$ compounds have been of great interest as these compounds are model systems for the investigation of an antiferromagnetic order caused by the bilinear exchange interactions of the trivalent rare earth ($R^{3+}$) ions in a strongly anisotropic magnetic system[10-18]. Among these, the $DyCu_2$ has gained much attention because it has been believed that the M – T graph exhibits distinct magnetic properties in different crystallographic directions[10]. The magnetization and the susceptibility measurements have unveiled two step magnetization process at 4.2 K for $DyCu_2$ single crystals[10]. Loewenhaupt *et. al.*, have reported the H – T phase diagram of $DyCu_2$ when the magnetic field parallel to a crystallographic a – axis, in which the $Dy^{3+}$ spins are oriented[16]. Detailed description and conversion of the magnetic Ising-axis in a high magnetic field parallel to the c - direction was found by Hashimoto *et. al*[15]. As for as the $DyCu_2$ properties are concerned, it consists of an orthorhombic structure and is known to exhibit an anti-ferromagnetic (AFM) phase below 27 K and shows magnetic field induced transitions at low temperature[18]. As this compound possesses the AFM to paramagnetic (PM) transition and a metamagnetic transition, large MCE has been successfully realized at 30 K on the bulk $DyCu_2$ compound by Arora *et. al*[9].

$DyCu_2$ is very ductile in nature as a result, it would be extremely difficult to reduce its dimension and study its physical properties. Hence, our first aim of this manuscript is to propose a method to prepare small flakes of the $DyCu_2$. As a result of the size reduction, physical properties may alter in a very impressive manner from their bulk counterpart, which might essentially be useful for various applications. Nevertheless, calculation of the MCE at reduced dimensions for the $DyCu_2$ has not been attempted until now, which is the main focus of the present work. In addition to the reduced size, correlation between the ferromagnetic (FM) and AFM phases which would arise due to the field induced transitions may alter the nature of magnetic transition. Such microscopic studies would be of great interest to understand the physics behind a strongly anisotropic $DyCu_2$ system in reduced form. Hence, in the second part of our work, we demonstrate our calculations pertinent to the MCE and the nature of magnetic transitions related to small flakes of $DyCu_2$.

As we stated above, our first aim is to prepare small flakes of $DyCu_2$. In order to attain them initially we prepared a polycrystalline $DyCu_2$ compound with the constituent elements (Dy of 99.9 wt% purity and Cu of 99.95 wt% purity) and employing in an arc furnace. The constituent elements were weighed in a stoichiometric proportion and were melted together under argon atmosphere. In order to attain homogeneous mixing, the melting was carried out several times. The weight loss after final melting was less than 0.5 %. We tried to prepare small flakes with the bulk $DyCu_2$ compound, however, it was very difficult to grind them as $DyCu_2$ is ductile in nature. Hence, these alloys were subsequently made in the form of ribbons by employing a single roller vacuum melt spinning unit under an argon atmosphere. The copper wheel speed over which the liquid metal quenched is maintained at constant speed of 34 m/sec. Surprisingly, the ribbons are very brittle in nature, while maintaining its structure. Subsequently small falkes (~ μm) of them were obtained using a planetary ball mill (Rectch, Germany), operating at a speed of 400 rpm in a medium of toluene for half an hour. Tungsten carbide balls of 2 mm diameter were used with a ball to powder ratio of 15:1. Phase purity of the sample was determined using the powder x – ray diffraction technique (Panalytical X-ray diffractometer) employing Cu - $K_\alpha$ radiation. Field emission Scanning electron microscope (FESEM - Zeiss) was employed to obtain back scattered electron images (BSE). Apart from the BSE images, we also have confirmed the composition and the nature of the microstructure in both the ribbons as well in the flakes by FESEM. A Dynacool VSM (Quantum design) was used to measure the magnetic field (H) *vs*. magnetization (M) and temperature (T) *vs*. magnetization (M) in the temperature range of 5 – 300 K and up to a maximum magnetic field of 50 kOe. Both the zero field cooling (ZFC) and field cooling (FC) behavior were mapped in M vs. T measurements.

Initially we discuss about the structural aspects of both the $DyCu_2$ ribbons and flakes. Subsequently we shall compare the magnetic properties of them with its bulk counterpart. Phase purity of the ribbons was checked with the powder x – ray diffraction technique (XRD). Fig. 1(a) and Fig. 1(b) reveal the XRD patterns of ribbon as well as flakes of $DyCu_2$ respectively. From fig. 1(a) it is evident that the most intense peak corresponds to (040), which essentially means that grains in melt – spun ribbon are oriented along (040) direction. Lattice parameter is calculated using least squares fitting and are found to be a = 4.312 (± 0.04) Å, b = 6.789 (± 0.067) Å and c = 7.240 (± 0.077) Å, which is consistent with the existing literature value[9]. Fig. 1 (b) shows the XRD pattern of the flakes of the $DyCu_2$ ribbons. It is apparent from the figure that reflections which are present are allowed reflections for an orthorhombic with the $CeCu_2$ type structure. It is evident from figure 1(b) that XRD peaks broadened after treating ribbons with ball milling, which is due to the reduced dimension. Calculated lattice parameter for the flakes are found to be a = 4.382 Å (± 0.043), b = 6.201 Å (± 0.059) and c = 7.375 Å (± 0.071). If we compare the lattice parameter of both the ribbons and flakes of $DyCu_2$, reasonable change is evident, hence, we believe that there exists distortion of the lattice. We also estimated the induced strain on flakes due to crystal imperfection and distortion using the Williamson-Hall method and it is found to be 1.5% (±0.29%). On top of that the preferential orientation towards (040) is absent for flakes of $DyCu_2$, which can be attributed to the various available orientations upon ball milling. Apart from the above, we also probed our sample for the microstructural details using the FE – SEM. Fig. 2(a) and 2(b) represents the BSE images of the ribbons and the flakes of $DyCu_2$. In both the cases, compositional analysis discloses that both the Dy and Cu are in proper composition in order to have 1:2 phase. BSE images infer that there exists no extra phase apart from the parent $DyCu_2$ phase. Inset of Fig. 2(b) shows the size distribution of the flakes which is fitted to Lorentz function.

Fig. 3(a) represents the susceptibility ($\chi$) vs. temperature (T) curve of the $DyCu_2$ ribbons at 1000 Oe in the temperature range 5 – 300 K. From 300 K, the susceptibility increases and peaks at 27 K, further decrease in temperature leads to decrease in the susceptibility, which is a typical nature of an antiferromagnetic material. Fig. 3(b) shows the $\chi$ vs. T curve for $DyCu_2$ flakes at 1000 Oe in the temperature range 5 – 300 K. From 300 K, the susceptibility increases until 5 K except a small peak at 27 K. The peak that is evident in both the ribbons and flakes at 27 K could be due to the AFM to a PM transition. It is worth mentioning that Arora *et. al*,[9] have reported that the $DyCu_2$ is a PM at room temperature but it becomes AFM below 27 K in its bulk form. If we clearly observe, the peak that is evident at 27 K is distinctly different for both the ribbons and flakes. In general, for a typical anti – ferromagnet, the magnetization should decrease below the Néel temperature, $T_N$. Same reflects in case of ribbons due to preferential orientation of grains toward (040) direction (Fig. 3(a)). However, as it is evident from Fig. 3(b) that the magnetization in both ZFC and FC increases which is unconventional for any AFM material. The reason for this kind of observation is that the M – T graph for $DyCu_2$ is different along different crystallographic directions[10]. Essentially there exists no signature of PM to AFM phase transition for the magnetization along b – axis which results increase in the magnetization continuously with decreasing the temperature. In contrast, the temperature gradient of the magnetization is very small along c – axis and the peak corresponding to the PM to AFM transition belongs to this axis. From the above discussion we strongly believe that the magnetization is not oriented along b – axis at least in case of ribbons. In addition, the peak intensity in Fig. 3(b) at transition is low if we compare the intensity that has been observed with 100 Oe[9]. The reduction in the intensity can be attributed to sensitivity of the peak for applied magnetic field. Insets of both Fig. 3(a) and Fig. 3(b) shows the inverse susceptibility ($1/\chi$) *vs.*T for 1000 Oe. At high temperatures, the gradient of the curve is obtained by fitting with a straight line, which essentially gave effective moment of 10.7 $\mu_B$. Extrapolation of the high temperature data to the temperature axis gave a paramagnetic Curie temperature ($\theta_p$) of 6.2 K. Both the above values are in accordance with the existing literature values[18, 19, 20].

From now onwards, we shall demonstrate the results pertinent to the MCE and nature of the magnetic transition of flakes of $DyCu_2$. Essentially we wanted to examine how the low dimensionality of $DyCu_2$ alters the MCE. Fig. 4 (a) & (b) shows the isothermal M vs. H for the flakes of $DyCu_2$ in the temperature range 5 - 95 K with a maximum applied magnetic field of 50 kOe. M *vs*. H behavior is distinctly different in both the phases (AFM & PM). Essentially, below the transition temperature 27 K, indeed there exists metamagnetic transitions, however, such transitions are absent in the high temperature M vs H graphs. Upon closer observation, below 27 K, magnetization behavior manifests a smooth field induced transition around 20 kOe from an AFM to a FM phase exactly resembles the one which has been reported by Sherwood *et. al.*[18]. Iwata *et. al.*[21], have reported the magnetic transition on the single crystal of $DyCu_2$ below 31.5 K. In their studies the argument of thought has been that demonstration of magnetization behavior in various directions of the single crystal. Particularly the magnetization along a - axis has been indicated as AFM to FM transition with a two-step process, meaning that the successive field – induced transitions from a commensurate AFM phase to an incommensurate AFM phase and then to a FM phase. In contrast, for the b - and c - axes the magnetization has increased monotonically at all the temperatures. One probable reason for the absence of a step like behavior in our results is the flakes of $DyCu_2$ are of polycrystalline in nature (which is evident from our XRD). Essentially, the aggregated magnetization behavior of three crystallographic axes is observed in the bulk magnetization results. We also estimated the ferromagnetic fraction by subtracting the magnetization arising from the antiferromagnetic component from the saturated magnetization values. Below the anti-ferromagnetic transition temperature, value of the ferromagnetic fraction is very low and is found to be ~10% for the magnetic field strength of 30 kOe and at 5 K. Percentage of the anti-ferromagnetism is dominant which could be due to the fact that the magnetic field is not sufficiently large enough to reach onset of the metamagnetic transition. This is also one of the reasons why we do see a strong anti-ferromagnetic behavior in M vs. T measurements.

Now we discuss the results pertinent to calculations of the MCE of flakes of $DyCu_2$ using magnetization data, which would in general results because of the coupling of the magnetic sublattice with the external magnetic field. Normally MCE can be realized by (a) change in magnetic entropy (b) heat capacity measurements. There exists three major contributions such as magnetic ($S_{magnetic}$), lattice ($S_{lattice}$) and electronic ($S_{electronic}$) for the total entropy of the system. Among those, the magnetic entropy change ($-\Delta S_M$) can be calculated using the Maxwell's equations. If we consider the $S_{lattice}$ and $S_{electronic}$ as magnetic field independent, magnetic contribution to the entropy change can be calculated using well known procedure[22-25]. The relation between temperature (T), magnetic field (H) and the magnetization of the material M to the MCE values, $\Delta S_M(T, \Delta H)$, is given by one of the Maxwell's relations[24]

$$\left(\frac{\partial S(T,H)}{\partial H}\right)_T = \left(\frac{\partial M(T,H)}{\partial T}\right)_H \qquad 1$$

By integrating above for an isothermal process

$$\Delta S_M(T,\Delta H) = \int_{H_1}^{H_2} \left(\frac{\partial M(T,H)}{\partial T}\right)_H dH \qquad 2$$

From the isothermal M vs H we obtain $\Delta S_M (T, \Delta H)$ by numerical integration of the equation 2. Using the above Maxwell's equation and isothermal M vs. H curves, the entropy change (($-\Delta S_M$) is estimated in the temperature range 5 – 95 K and up to the magnetic field strength of 50 kOe. Variation of the $-\Delta S_M$ at different magnetic fields is shown in Fig. 5. Owing to the increase in the magnetic fields up to 50 kOe, the change in -$\Delta S_M$ of 4.31 J/kg – K is observed at an AFM – PM transition (27 K). A maximum value of -6.9 J/kg-K at 27.5 K has been reported for bulk $DyCu_2$ at the magnetic field strength of 45 kOe[9]. Sampathkumaran *et. al.*, have reported a maximum $\Delta S_M$ of – 11.8 J/kg-K near the transition and for a magnetic field strength of 55 kOe on an antiferromagnetic compound $Gd_2PdSi_3$[26]. $\Delta S_M$ of -19 J/kg – K close to 30 K in 50 kOe magnetic field has been reported on DyNiAl compound[27]. $\Delta S_M$ of -2.8 J/kg-K has been reported in MnCoGe ribbons for a magnetic field change of 50 kOe[28]. In a sharp contrast, yet in another study $\Delta S_M$ of -30 J/kg – K has been reported in the melt-spun ribbons of $Ni_{52}Mn_{26}Ga_{22}$ for a magnetic field change of 50 kOe[29]. $\Delta S_M$ ~ 6.0 J/kg-K and ~1.6 J/kg-K for a field change of 20 kOe has been observed in the $Ni_{49}Mn_{37.4}Sn_{13.6}$ and $Ni_{50}Mn_{34.5}Sn_{15.5}$ ribbons, respectively[30]. Decrease in maximum value for $\Delta S_M$ in flakes of $DyCu_2$ could be due to the decrease in the dimension. From the inset of Fig. 5 ($\Delta S_M$ vs. T at 50 kOe) it is also worth noting that the $\Delta S_M$ peak found to be strongly dependent on the dimension of $DyCu_2$. Essentially $\Delta S_M$ peak broadened marginally compared with the bulk $DyCu_2$[9]. This can be attributed to the structural disorder which could be due to the significant increase in total grain boundary volume.

Utility of the magnetocaloric material cannot be decided not only based on the magnitude of the $\Delta S_M$ but also based on its temperature dependence[31-32]. Essentially, the refrigeration capacity (RC) is widely used to approximate the amount of thermal energy that can be transferred by the magnetic refrigerant between the cold and hot sinks in one ideal thermodynamic cycle. In approximation, the RC is described by $\Delta S_M$ x $\delta T_{FWHM}$, where $\delta T_{FWHM}$ is full width at half maximum (FWHM) of $\Delta S_M$ (T) peak.

Refrigeration capacity (RC) is calculated using the equation

$$RC = \int_{T_{cold}}^{T_{hot}} \Delta S_M dT$$

Inset of Fig. 5 represents the $\Delta S_M$ vs. T at 50 kOe. Estimated FWHM is used to calculate RC using the above equation ($T_{cold}$ = 5 K and $T_{hot}$ = 47 K). RC at the transition is computed and it is found to be 165.84 J/kg at 50 kOe, almost comparable with the value reported on $DyCu_2$ bulk sample (193.34 J/kg) and more in comparison with the one reported for the melt spun ribbons of $Ni_{52}Mn_{26}Ga_{22}$ (75 J/kg at 50 kOe)[29]. RC value of 24 J/kg has been observed for $Ni_{49}Mn_{37.4}Sn_{13.6}$ at 20 kOe[30]. In contrast RC value of 89 J/kg has been reported for $Mn_{50}Ni_{40}In_{10}$ preferentially textured melt spun ribbons. Marginal decrease in RC value when compared with the bulk $DyCu_2$ samples could be due to the compressive strain as a result of the ball milling. At the same time one cannot also rule out the decrease in magnetization upon the size reduction. RC found in this material is quite large if you compare with other melt-spun ribbons. At temperatures higher than $T_N$ large value of $\Delta S_M$ is obtained hinting that large temperature window where essentially the flakes of $DyCu_2$ may be used as potential candidate as magnetic refrigerant material.

We also have studied the nature of magnetic transition using the Arrott plots ($M^2$ vs. H/M). Apart from the determination of the Curie temperature ($T_C$) in the ferromagnetic materials, one can also utilize Arrott plots to estimate Nèel temperature ($T_N$) of an anti-ferromagnetic material[33-36]. Fig. 6 shows the Arrott plots of $DyCu_2$ flakes in the temperature range of 5 – 90 K. From these graphs it is evident that the isotherms are nonlinear and the curvature at low $M^2$ values is distinctly different above and below the transition temperature (27 K). Strictly speaking, a negative curvature is apparent below 27 K where as a positive curvature is evident above 27 K. At high magnetic fields $M^2$ varies linearly with respect to the magnetic field. Such a deviation from the linearity of isotherms around origin has been observed in the spin – glass systems like $Y_{1-x}Fe_x$[37], $Gd_{37}Al_{63}$[38], $Al_{37}Mn_{30}Si_{33}$[39]. It is worth noting that similar kind of behavior has been observed by Rohit *et. al*, on $Ni_{48}Co_6Mn_{26}Al_{20}$[36] the polycrystalline ribbons. In addition, from the Arrott plots, we could not able to estimate transition temperature since the extensions of the plots do not reach the origin of the $M^2$ vs. H/M graph. Hence it could be possible that the magnetic transition is broadly spread both in temperature and the magnetic field due to a competition between the FM and the AFM phases.

According to Aharony & Pytte[40], deviation of $M^2$ from linear behaviour in Arrott plots is due to random anisotropy or random field presented in the system. Essentially, if we have a system with mixed phases (FM and AFM), the impurity spins (FM) can couple with host phase (AFM)[41]. Under such conditions, if we apply a finite magnetic field, it will have a reversed sign on the impurity spins which can lead to a random component of the field.

On the other hand, in a system with spins, if there is a contribution from orbital moment to total magnetic moment, adjacent local magnetic moments would be influenced by local crystal fields in addition to exchange interactions. Since the $DyCu_2$ is made up of Dy and Cu, the local crystal field of Dy depends on the environment that would be created by the neighbouring atoms (Cu). This may create random anisotropy in the system. Detailed analysis on random field and random anisotropy indeed require further theoretical investigation.

In conclusion, we have successfully demonstrated our efforts in preparing flakes of $DyCu_2$. On top of that, we also have demonstrated the magnetocaloric properties and the nature of magnetic transition in $DyCu_2$ flakes. The value of -$\Delta S_M$ in $DyCu_2$ flakes decreased from 6.91 J/kg – K to 4.31 J/kg – K, which could be due to the low dimensionality of $DyCu_2$. Indeed, there exists a strong dependence of the $\Delta S_M$ peak on the dimension. $\Delta S_M$ peak broadened marginally compared with the bulk $DyCu_2$, which is attributed to significant increase in total grain boundary volume. There exists marginal decrease in RC value when compared with bulk $DyCu_2$, which can be attributed to the decrease in magnetization as a result of strain that we applied using ball milling. As these flakes consists larger $\Delta S_M$ values above $T_N$, larger temperature window is available for magnetic refrigeration where essentially flakes of $DyCu_2$ may be used as potential candidate as the magnetic refrigerant material.

We thank Indian Institute of Technology Hyderabad for the financial support. SNJ would also like to thank Department of Science and Technology (DST) (Project #SR/FTP/PS-190/2012) for financial support.

**Figure captions:**

**Fig. 1**: Powder x – ray diffraction of (a) DyCu$_2$ ribbon (b) Flakes of DyCu$_2$. It is evident that in case of ribbon, the preferential direction is towards (040). However, such preferential direction is absent in case of flakes due to polycrystalline in nature.

**Fig. 2:** Back scattered electron image (BSE) of DyCu$_2$ ribbons and flakes. From the above images it is clear that there exists no extra phase apart from parent 1:2 phase.

**Fig. 3**: Susceptibility χ vs. T graph for (a) DyCu$_2$ ribbon (b) DyCu$_2$ flakes. Insets reveal inverse susceptibility 1/χ vs. T graph for ribbons and flakes of DyCu$_2$ respectively.

**Fig. 4**: Isothermal Magnetic field H vs. Magnetization M for flakes of DyCu$_2$. (a) M vs. H graphs from 5 – 50 K. Arrow indicates metamagnetic transitions due to the application of magnetic field. (b) M vs. H graphs from 55 – 95 K. Metamagnetic transitions are absent at high temperatures.

**Fig. 5:** -ΔS (J/kg – K) vs. T (K) for the flakes of $DyCu_2$. It is evident that graphs peaks at 27 K which is a transition temperature from anti-ferromagnetic to paramagnetic phase. Inset shows the $-\Delta S_M$ vs. T at 50 kOe. Calculated FWHM used to estimate RC values.

**Fig. 6:** $M^2$ vs. (H/M) plots for the flakes of $DyCu_2$.

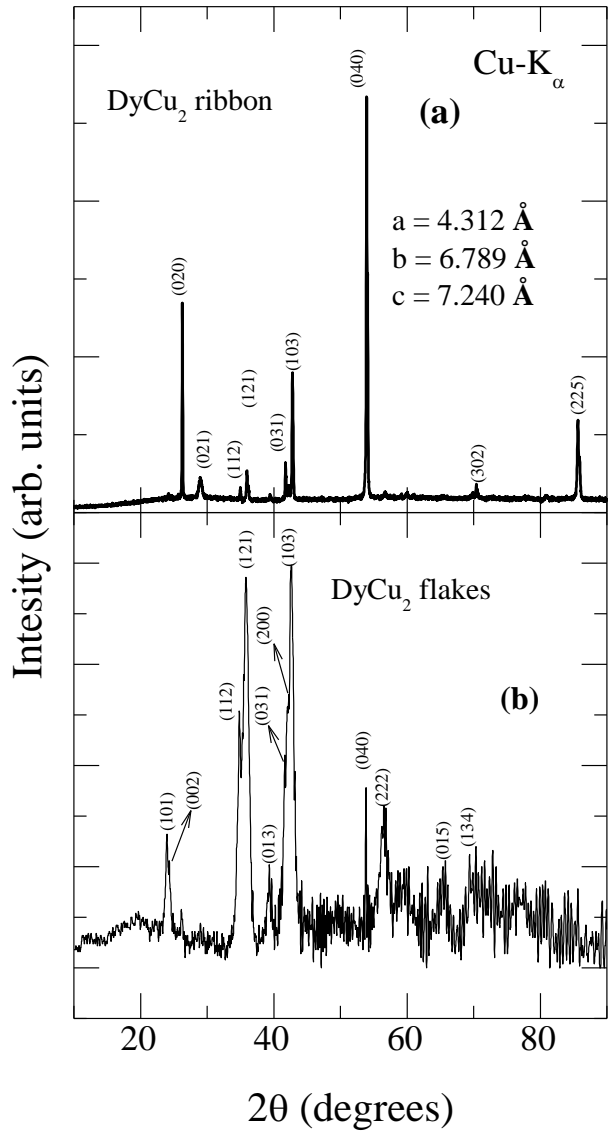

**Fig. 1**: Powder x − ray diffraction of (a) DyCu$_2$ ribbon (b) Flakes of DyCu$_2$. It is evident that in case of ribbon, the preferential direction is towards (040). However, such preferential direction is absent in case of flakes due to polycrystalline in nature.

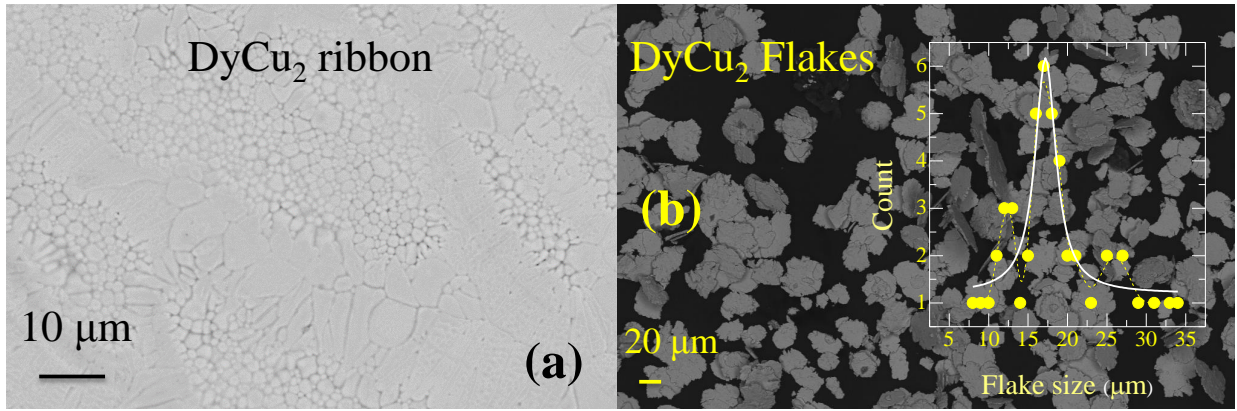

**Fig. 2:** Back scattered electron image (BSE) of DyCu$_2$ ribbons and flakes. From the above images it is clear that there exists no extra phase apart from parent 1:2 phase. Inset of (b) depicts the size distribution of flakes.

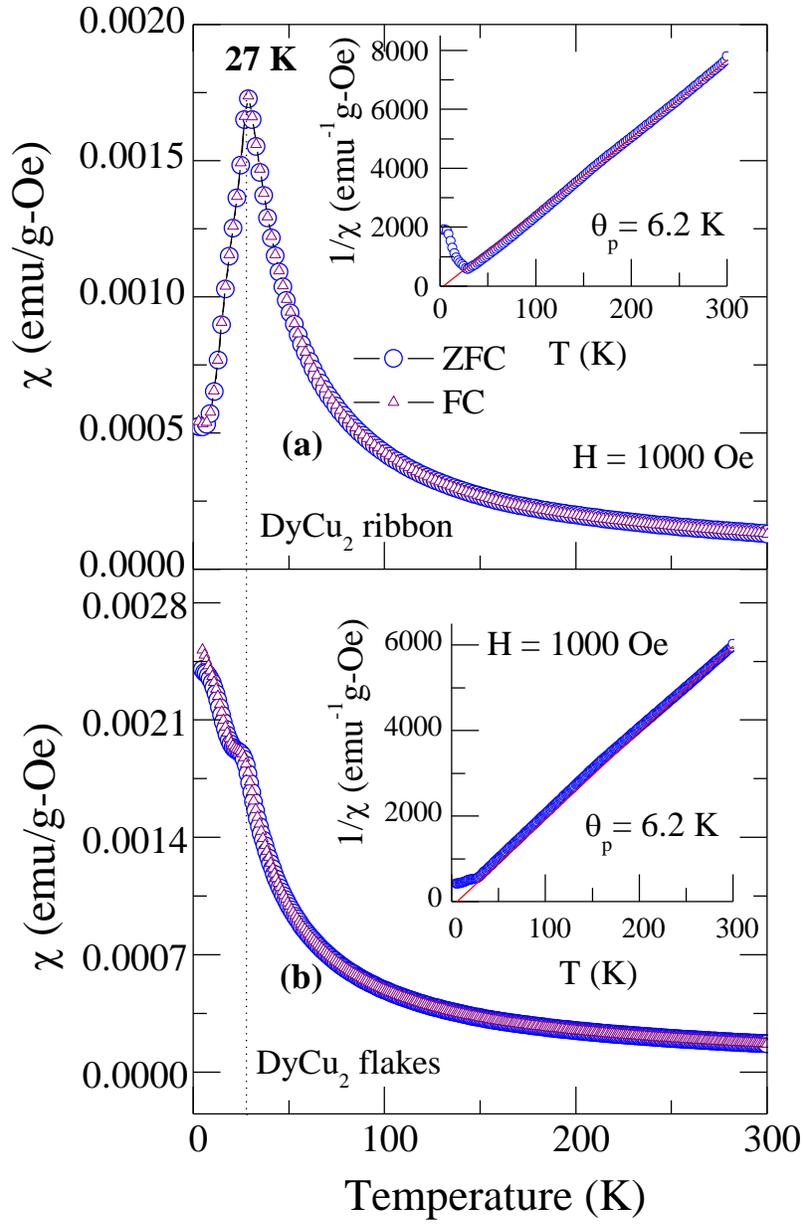

**Fig. 3**: Susceptibility χ vs. T graph for (a) DyCu$_2$ ribbon (b) DyCu$_2$ flakes. Insets reveal inverse susceptibility 1/χ vs. T graph for the ribbons and the flakes of DyCu$_2$ respectively.

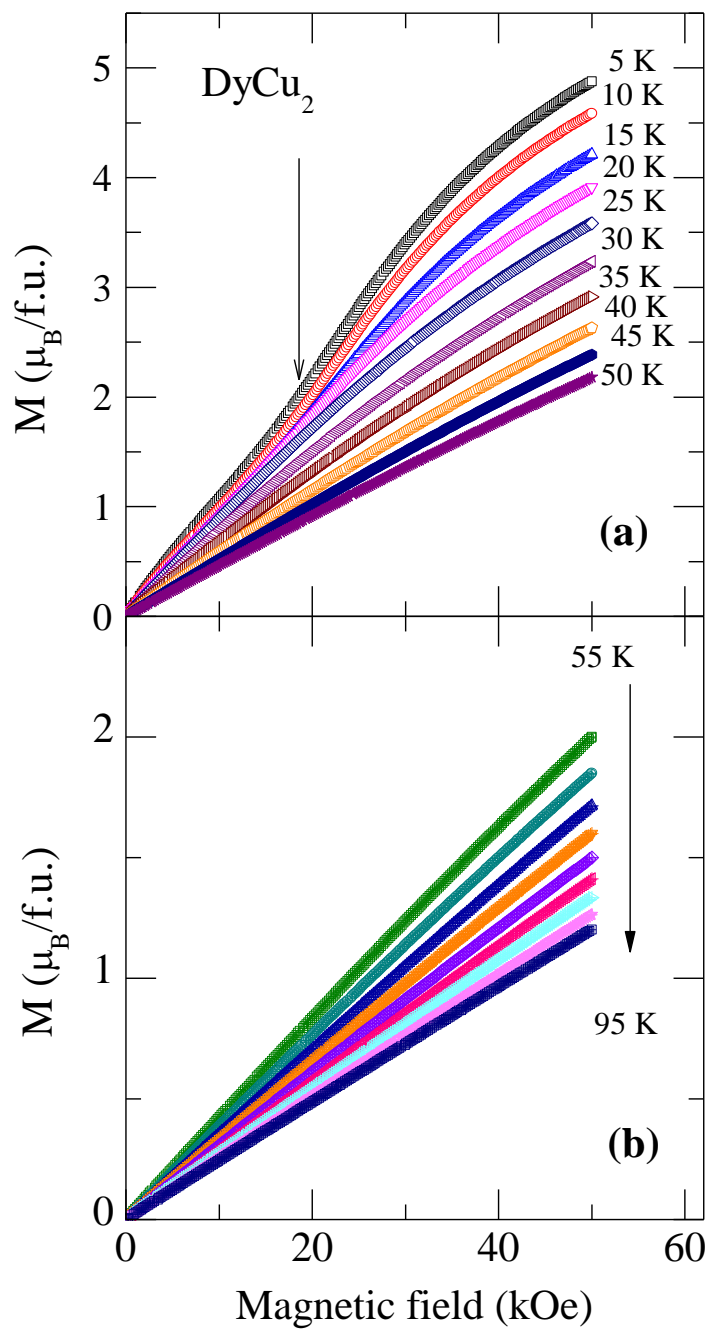

**Fig. 4**: Isothermal Magnetic field H vs. Magnetization M for flakes of $DyCu_2$. (a) M vs. H graphs from 5 – 50 K. Arrow indicates metamagnetic transitions due to the application of magnetic field. (b) M vs. H graphs from 55 – 95 K. Metamagnetic transitions are absent at high temperatures.

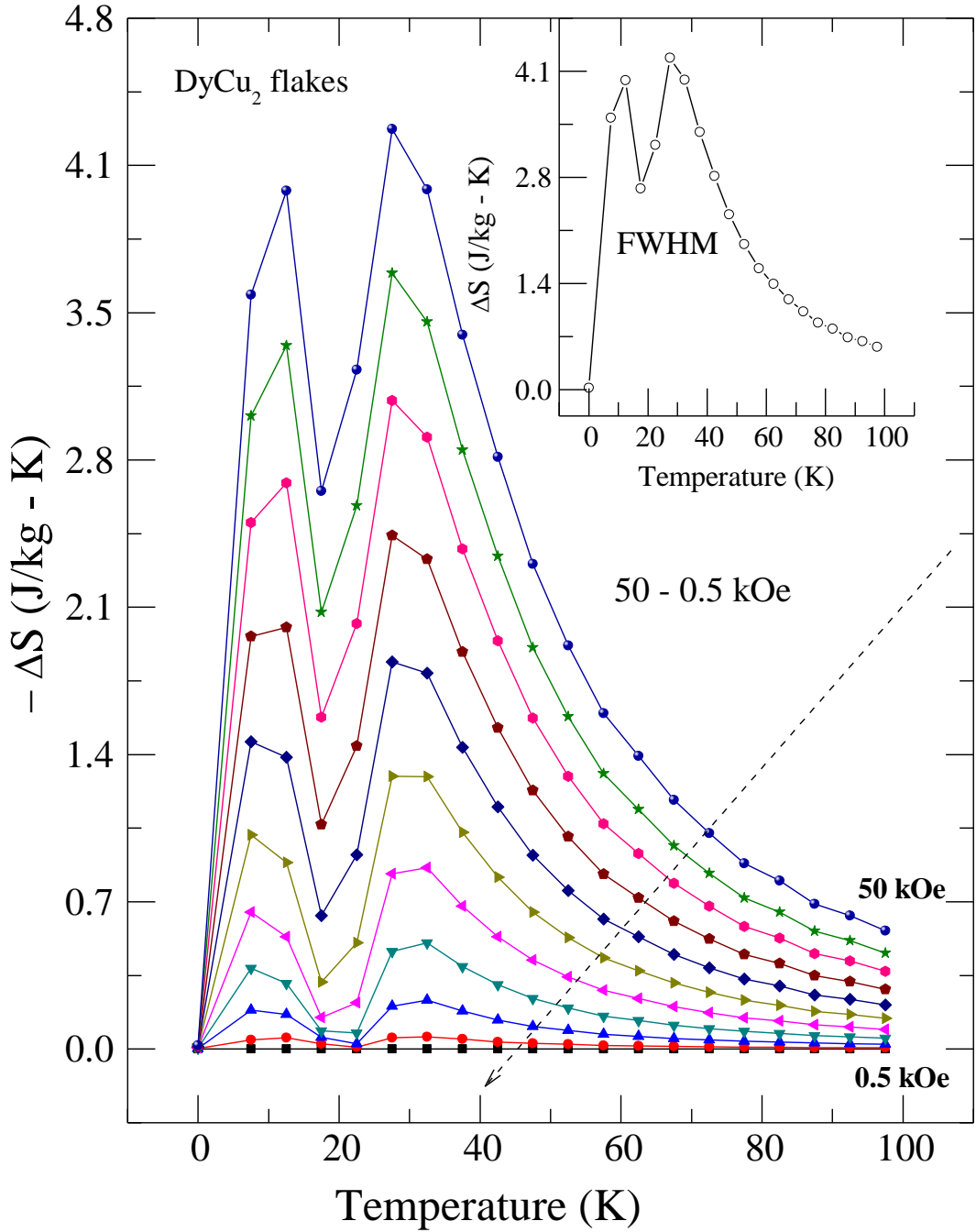

**Fig. 5:** -ΔS (J/kg – K) vs. T (K) for the flakes of DyCu$_2$. It is evident that graphs peaks at 27 K which is a transition temperature from anti-ferromagnetic to paramagnetic phase. Inset shows the −ΔS vs. T at 50 kOe. Calculated FWHM used to estimate RC values.

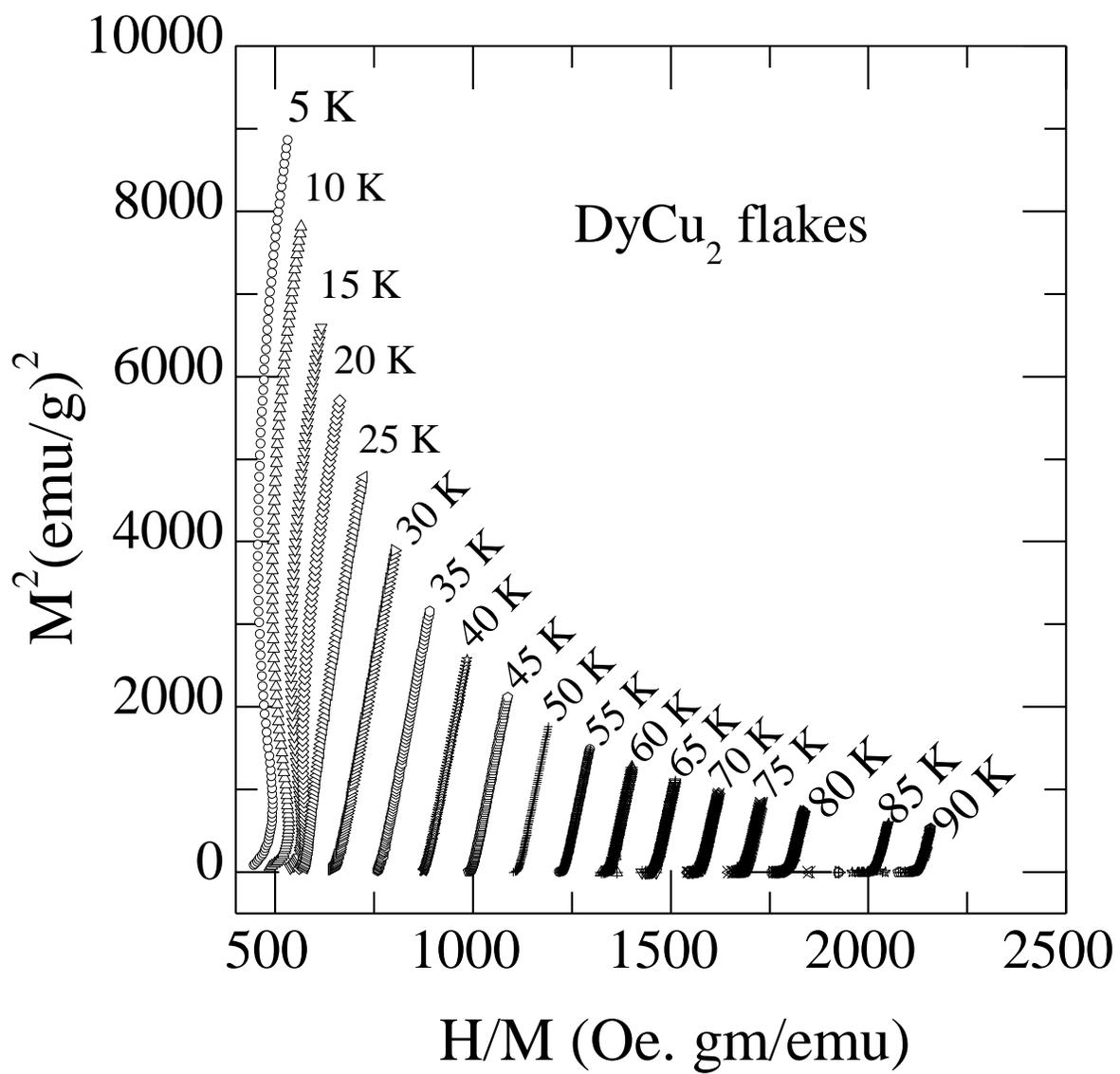

**Fig. 6:** $M^2$ vs. (H/M) plots for the flakes of $DyCu_2$.